\documentclass[aps,prl,twocolumn,superscriptaddress,showpacs,floatfix,reprint]{revtex4-1}
\usepackage{makeidx}
\usepackage{amsfonts}
\usepackage{amssymb}
\usepackage{enumerate}
\usepackage{latexsym}
\usepackage{bbm}
\usepackage{graphicx}
\usepackage{float}

\newcommand{\beq}{\begin{equation}}
\newcommand{\eeq}{\end{equation}}
\newcommand{\beqa}{\begin{eqnarray}}
\newcommand{\eeqa}{\end{eqnarray}}
\newcommand{\refeq}[1]{Eq.~(\ref{eq:#1})}
\newcommand{\reffig}[1]{Fig.~\ref{fig:#1}}

\newcommand{\ket}[1]{\left\vert #1 \right\rangle}
\newcommand{\bra}[1]{\left\langle #1 \right\vert}

\newcommand{\psc}{\mathbb{P}_1^{\mathrm{sc}}}

\begin{document}

\title{Experimental determination of the degree of quantum polarisation of continuous variable states}

\author{Christian Kothe}
\affiliation{Department of Physics, Technical University of Denmark, Fysikvej, DK-2800 Kongens Lyngby, Denmark}
\author{Lars Skovgaard Madsen}
\affiliation{Department of Physics, Technical University of Denmark, Fysikvej, DK-2800 Kongens Lyngby, Denmark}
\author{Ulrik Lund Andersen}
\affiliation{Department of Physics, Technical University of Denmark, Fysikvej, DK-2800 Kongens Lyngby, Denmark}
\author{Gunnar Bj\"{o}rk}
\affiliation{Department of Applied Physics, Royal Institute of Technology (KTH), AlbaNova University Center, SE-106 91 Stockholm, Sweden}
\date{\today}

\begin{abstract}

We demonstrate excitation-manifold resolved polarisation characterisation of continuous-variable (CV) quantum states. In contrast to traditional characterisation of polarisation that is based on the Stokes parameters, we experimentally determine the Stokes vector of each excitation manifold separately. Only for states with a given photon number does the methods coincide. For states with an indeterminate photon number, for example Gaussian states, the employed method gives a richer and more accurate description. We apply the method both in theory and in experiment to some common states to demonstrate its advantages.

\end{abstract}


\maketitle

Polarisation is one of the key parameters of the electro-magnetic field as demonstrated by the plethora of different applications. To mention a few, the classical polarisation is used in thin-film ellipsometry \cite{azzam}, near-field microscopy \cite{werner97}, remote sensing \cite{schott2009} and light scattering \cite{barron}. In recent years, the concept of polarisation has also found a footing in quantum optics and in quantum information science where the information is efficiently encoded in the polarisation degree of freedom. This has lead to the demonstrations of polarisation entanglement \cite{kwiat95}, teleportation of the quantum polarisation \cite{bouwmeester97} and quantum key distribution based on quantum polarisation encoding \cite{muller, bennett92}. Due to the importance of quantum polarisation in these applications and others, it is important to be able to quantify the degree of quantum polarisation.

Classically, the degree of polarisation is a simple expression of the mean values of the Stokes parameters \cite{stokes} which can be straightforwardly measured \cite{bowen}. It has been suggested to use a similar expression in the quantum domain as a measure of the degree of quantum polarisation \cite{fano, collett, klyshko}. However, it was soon realised that this polarisation parameter is insufficient to characterise the degree of polarisation for many quantum states since it classifies some states as being unpolarised although they are polarised and vice versa (see for example \cite{usachev} and references therein). This inconsistency calls for a new measure that more accurately characterises the polarisation of quantum states.

Several attempts have been made to quantify the degree of quantum polarisation differently (see \cite{bjoerk10} for an overview), the most prominent ones being the distance-based \cite{ghiu,klimov05} or Q-function based measures \cite{luis02, *luis05}. While they all fully satisfy the requirements for a polarisation measure their complexity makes them extremely hard to access in a time-efficient manner.

In this Letter, we suggest a new and simple measure of quantum polarisation and implement it experimentally. However, in contrast to the semi-classical measure, the new measure accounts for the polarisation in each excitation manifold which leads to a better characterisation of the quantum polarisation. 

\textit{Polarisation measures} - The polarisation of a classical, electro-magnetic field is uniquely described by the Stokes parameters which can be written as
\beqa
\nonumber
S_0=\left|a_1\right|^2+\left|a_2\right|^2&, &S_1=a_H^\ast a_V+a_H a_V^\ast,\\
\label{eq:stokesclassic}
S_2=-i\left(a_H^\ast a_V-a_H a_V^\ast\right)&, &S_3=\left| a_H\right|^2-\left| a_V\right|^2,
\eeqa
where $a_{H}$ and $a_{V}$ denote the amplitudes of the field in two linearly polarised orthogonal modes $H$ and $V$. From these, the classical degree of polarisation $\mathbb{P}^{\mathrm{cl}}$ is defined as
\beq
\mathbb{P}^{\mathrm{cl}}=\frac{\sqrt{S_1^2+S_2^2+S_3^2}}{S_0}.
\eeq

The degree of quantum polarisation has previously been defined as a direct translation of the classical degree \cite{fano, collett, klyshko}:
\beq
\psc\left(\hat{\rho}\right)=\frac{\sqrt{\langle\hat{S}_1\rangle^2+\langle\hat{S}_2\rangle^2+\langle\hat{S}_3\rangle^2}}{\langle\hat{S}_0\rangle},
\label{eq:psc}
\eeq
where $\hat{\rho}$ is the quantum state under scrutiny \cite{firstorder} and the Stokes operators ($\hat S_0, \hat S_1, \hat S_2, \hat S_3$) are found by the canonical quantisation of the field amplitudes in the expressions~(\ref{eq:stokesclassic}). As (\ref{eq:psc}) is undefined for the two-mode vacuum state, the degree needs the supplementary definition $\psc(\ket{0}_H\ket{0}_V)=0$ as this state is invariant under any polarisation transformation and is therefore unpolarised \cite{prakash71, *prakash74}.

Despite the seemingly correct translation from classical to quantum polarisation, the definition in (\ref{eq:psc}) yields an inconsistent quantification of the degree of polarisation~\cite{klimov10}. This can be illustrated by some simple examples: According to the definition in (\ref{eq:psc}), the state $\ket{\Psi}_H \ket{0}_V$  is fully polarised (that is $\psc=1$) for any pure state $\ket{\Psi}_H\neq \ket{0}$. This implies the unpalatable consequence that states arbitrarily close to the two-mode vacuum are fully polarised and renders the measure discontinuous as a function of the state excitation. As another example of its failure, we consider the state $\ket{\Xi}=\frac{1}{\sqrt{3}}\left(\sqrt{2}\ket{1,0}_{H,V}+e^{i\varphi}\ket{0,2}_{H,V}\right)$
where $\varphi\in\left[0;2\pi\right)$. Applying the semi-classical polarisation (\ref{eq:psc}) yields $\psc(\ket{\Xi})=0$ and thus implies that $\ket{\Xi}$ is unpolarised. This means that the state should be polarisation transformation invariant~\cite{prakash71, *prakash74}. However, $\ket{\Xi}$ is not polarisation transformation invariant: Under, a $\pi/2$ polarisation rotation of the state, the state is transformed into  $\frac{1}{\sqrt{3}}\left(\sqrt{2}\ket{0,1}_{H,V}-e^{i\varphi}\ket{2,0}_{H,V}\right)$, which is orthogonal to $\ket{\Xi}$. Thus $\ket{\Xi}$ is not invariant under polarisation transformation and it is therefore clear that the definition in (\ref{eq:psc}) falls short in quantifying the degree of quantum polarisation. We note that a common property of the aforementioned examples is that the photon number $n$ is not a fixed quantity.

From the above discussion, it is clear that the semi-classical definition is unsuitable for determining the degree of polarisation for many quantum states. As the main result of this paper, we propose a new definition of the degree of polarisation that circumvents the shortcomings of the previous definition:
\beq
\mathbb{P}_1\left(\hat{\rho}\right)=\sum\limits_{N=1}^{\infty}p_{N}\frac{\sqrt{\langle\hat{S}_{1,N}\rangle^2+\langle\hat{S}_{2,N}\rangle^2+\langle\hat{S}_{3,N}\rangle^2}}{\langle\hat{S}_{0,N}\rangle},
\label{eq:new degree} \eeq
where $p_{N}= \textrm{Tr}(\hat{\openone}_N \hat{\rho})$, $\hat{\rho}_N = (\hat{\openone}_N \hat{\rho} \hat{\openone}_N)/p_N$, $\langle\hat{S}_{j,N}\rangle = \textrm{Tr}(\hat{S}_j \hat{\rho}_N)$, and $\hat{\openone}_N = \sum_{m=0}^{N}\ket{m,N-m} \bra{m,N-m}$, so that $\hat{\rho}_{N}$ is the normalised $N$-photon projection of the state's density matrix. The polarisation degree is quantified by a weighted sum of the semi-classical degree of polarisation in each excitation manifold of the state (except for $N=0$). In other words, every excitation manifold is treated separately. As is clear from the definition, $\mathbb{P}_1$ coincides with $\psc$ when the number of photons is a fixed quantity. The two definitions also become approximately equal for classical-like states such as coherent states with $\langle\hat{S}_0\rangle\gg 1$. For many other states the two definitions give different results, and in the following we argue that $\mathbb{P}_1$ gives a better assessment of the polarisation properties of quantum states than $\psc$. For example, we find the intuitively correct results  $\mathbb{P}_1\left(\ket{\Xi}\right)=1$ and $\mathbb{P}_1\left(\ket{\Psi}_H \ket{0}_V\right)\neq 1$. Below, we will examine, theoretically and experimentally, some properties of $\mathbb{P}_1$ for continuous variable Gaussian states, and we will in particular focus on moderately excited coherent and squeezed states \cite{andersen10}.

\begin{figure}
	\centering
		\includegraphics[scale=.5]{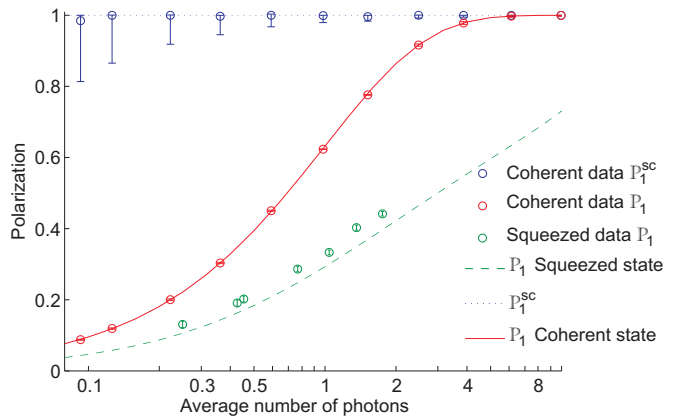}
		\caption{Comparison between $\mathbb{P}_1$ and $\psc$ as a function of the average photon number for the states $\ket{\Psi(\alpha)}_H\ket{0}_V$ and $\ket{\psi(\xi(r))}_H\ket{0}_V$ as defined in the text. The lines represent theoretical predictions while the circles indicate experimental values.
		}
	\label{fig:Ponemodestate}
\end{figure}

First we consider a two-mode state in which one of two orthogonal polarisation modes is vacuum whereas the other one is a coherent state; $\ket{\Psi(\alpha)}=\ket{\alpha}_{H}\otimes\ket{0}_{V}$, where $\alpha=a e^{i\phi}$ is the complex amplitude of the coherent state ($a\in\mathbb{R}^{+}_{0}$, $\phi\in\left[0;2\pi\right)$).
For this state, the semi-classical degree of polarisation is unity, $\psc\left(\ket{\Psi(\alpha)}\right)=1$, for all values of $\alpha$ except $\alpha =0$. On the other hand, using the new measure we find
\beq
\mathbb{P}_1\left(\ket{\Psi(\alpha)}\right)=1-e^{-\left|\alpha\right|^2}.
\label{eq:coherentstatedegofpol}
\eeq
which is continuous $\forall \ \alpha$; $\mathbb{P}_1\left(\ket{\Psi(\alpha)}\right)\rightarrow 0$ when $\alpha\rightarrow 0$, and for large amplitudes, $\mathbb{P}_1\left(\ket{\Psi(\alpha)}\right)\rightarrow 1$ when $|\alpha| = a \gg 1.$. Therefore, the classical and quantum  limits, respectively corresponding to large and small amplitudes, are smoothly connected. $\mathbb{P}_1\left(\ket{\Psi(\alpha)}\right)$ is illustrated in \reffig{Ponemodestate} by the solid line, $\psc\left(\hat{\rho}\right)$ is shown by the dotted line. Eq. (\ref{eq:coherentstatedegofpol}) can be easily generalised to any two-mode coherent states,  $\ket{\alpha}_H\ket{\beta}_V$, which after an appropriate transformation can be written as a one mode coherent state, $\ket{\alpha'}\ket{0}$ in some other polarisation basis with $|\alpha'|^2=|\alpha|^2 + |\beta|^2$.
The degree of quantum polarisation of any two-mode coherent state is therefore given by \refeq{coherentstatedegofpol} with $|\alpha|^2 \rightarrow |\alpha|^2 + |\beta|^2$.

Next we consider the degree of quantum polarisation for single-mode squeezed states; $\ket{\varphi(\xi,\alpha)}=\hat{D}(\alpha)\hat{\mathcal{S}}(\xi)\ket{0}_H\otimes \ket{0}_V$ where $\hat{D}$ is the displacement operator and $\hat{\mathcal{S}}(\xi)=\exp\left[(\xi^{\ast}\hat{a}^2-\xi\hat{a}^{\dagger 2})/2\right]$ is the squeezing operator with $\xi=r e^{i\theta}$, $r\in\mathbb{R}^{+}_{0}$, $\theta\in\left[0;2\pi\right)$ being the squeezing parameter. For this state we find $\psc(\ket{\varphi(\xi,\alpha)})=1$, whereas
\beqa
&\mathbb{P}_1&(\ket{\varphi(\xi,\alpha)})=
\label{eq:onemodegeneral}\\
\nonumber
&1&-\frac{1}{\cosh(r)}\exp\left[-|\alpha|^2-\frac{1}{2}\left(\alpha^{\ast 2} e^{i\theta}+\alpha^2 e^{-i\theta}\right)\tanh(r)\right].
\eeqa
which is illustrated in \reffig{Ponemodestate} as a function of the average number of photons for a squeezed vacuum state (that is, $\alpha =0$ and $\langle n_{sqz}\rangle = \sinh(r)$).
We clearly see that $\mathbb{P}_1$ differs significantly from $\psc$ for basically all practical squeezing values. We also note that the degree of quantum polarisation (according to the new measure) is not solely determined by the photons associated with the coherent excitation but also by the photons responsible for the squeezing. However, for a fixed number of photons the coherent state is significantly more polarised than a squeezed vacuum state. Finally, we note that for the generalised squeezed state in (\ref{eq:onemodegeneral}), the degree of polarisation also depends on the squeezing angle, $\theta$, relative to the phase of the displacement, $\phi$: It is maximised for amplitude squeezing ($\theta-2\phi=0$) and minimised for phase squeezing ($\theta-2\phi=\pi/2$).

\begin{figure}
		\begin{tabbing}
		\includegraphics[scale=.4]{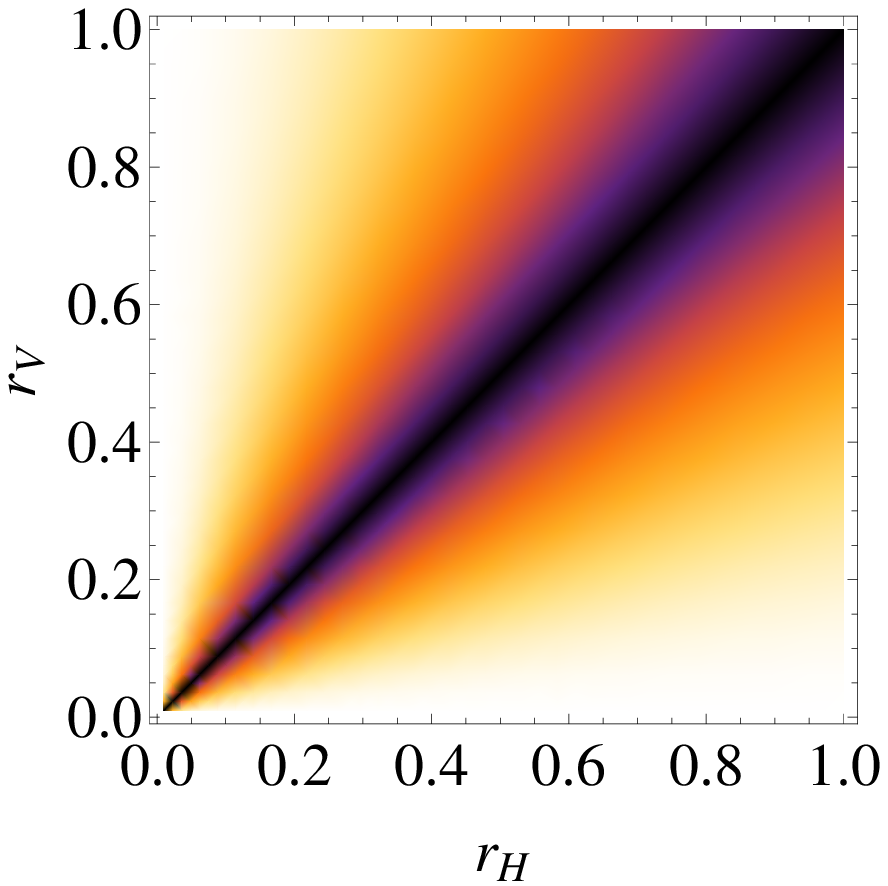}\=
		\includegraphics[scale=.4]{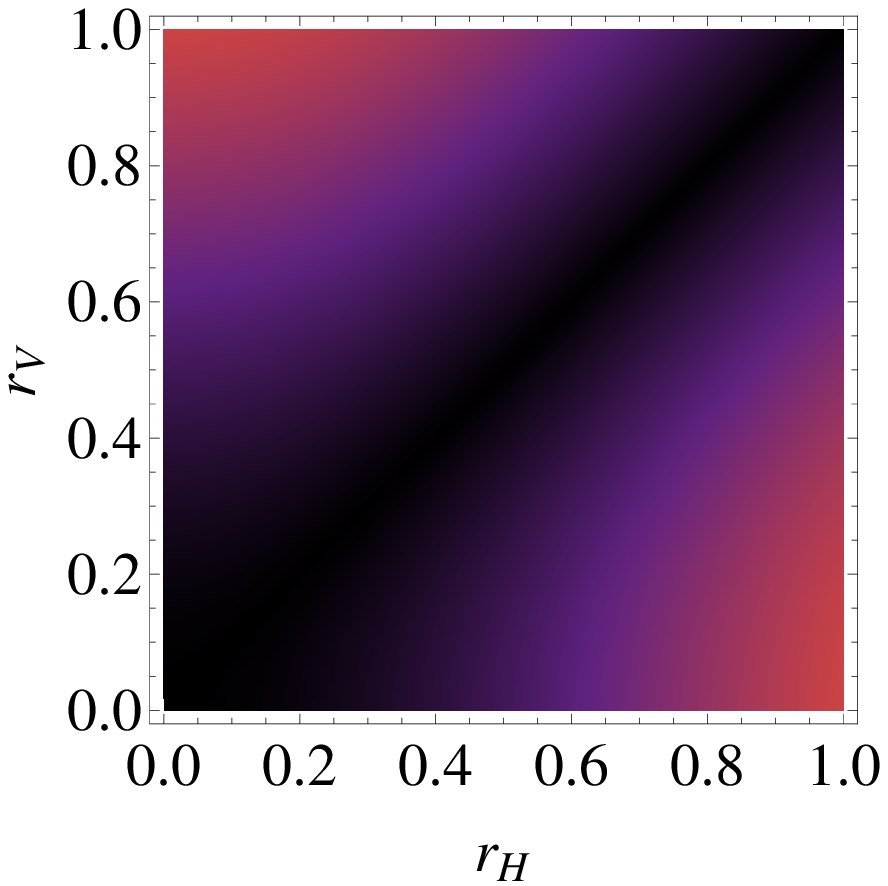}\=\\
		\includegraphics[scale=.4]{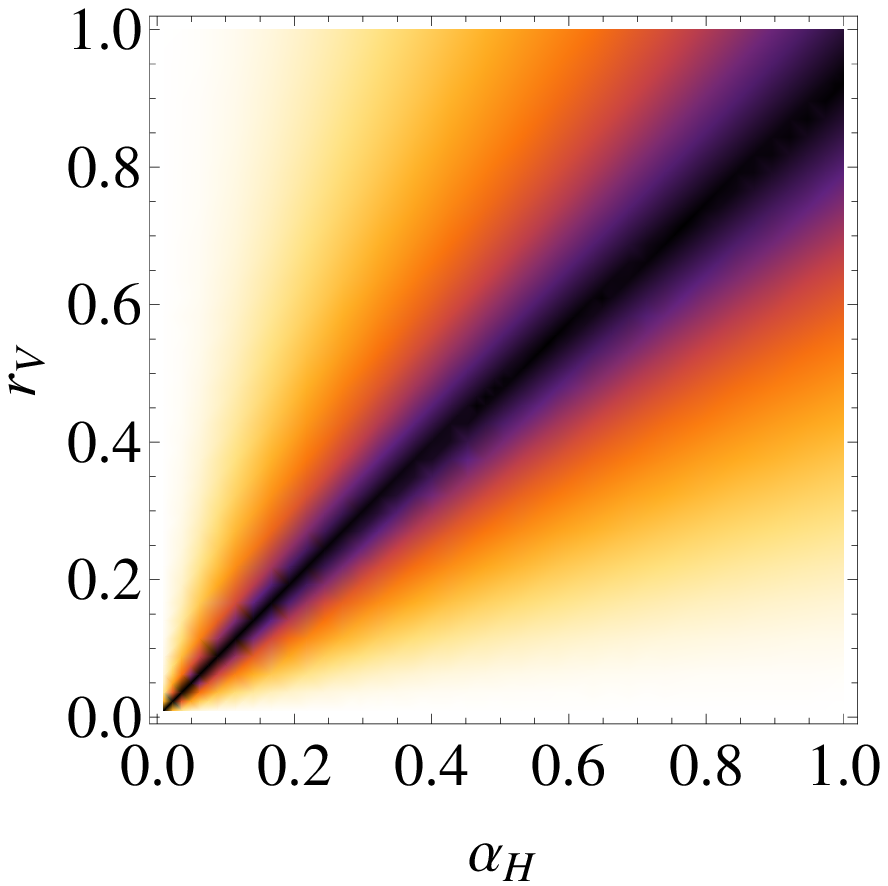}\>
		\includegraphics[scale=.4]{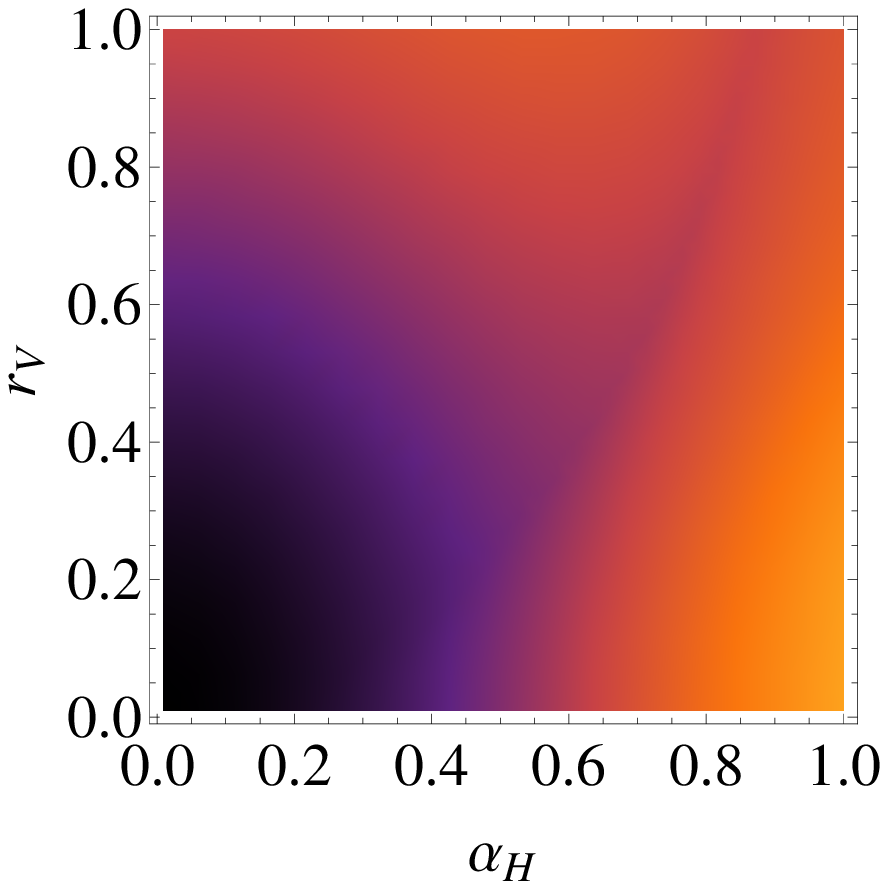}\>\\
		\includegraphics[scale=.4]{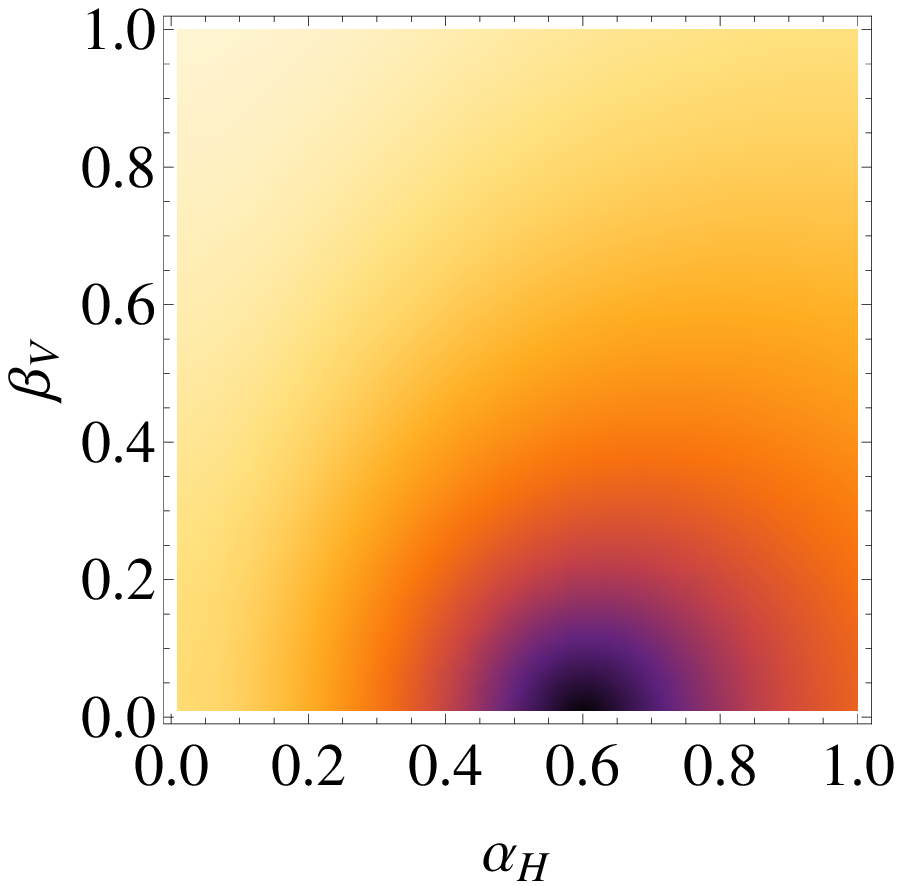}\>
		\includegraphics[scale=.4]{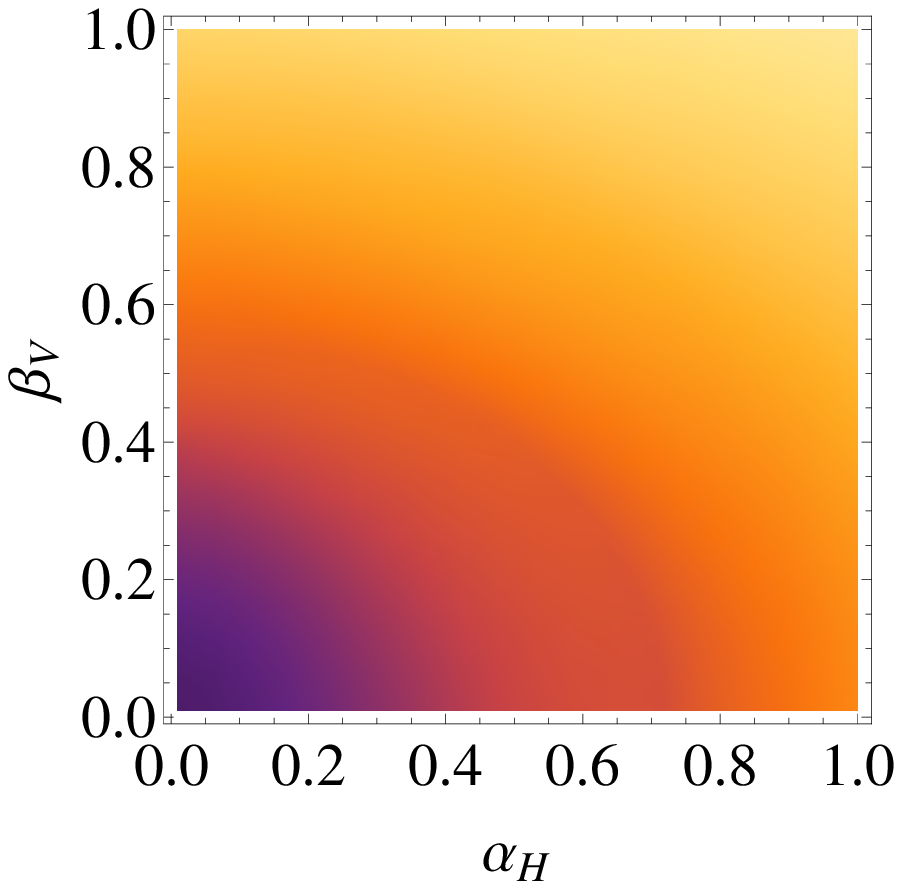}\>
		\includegraphics[scale=.3]{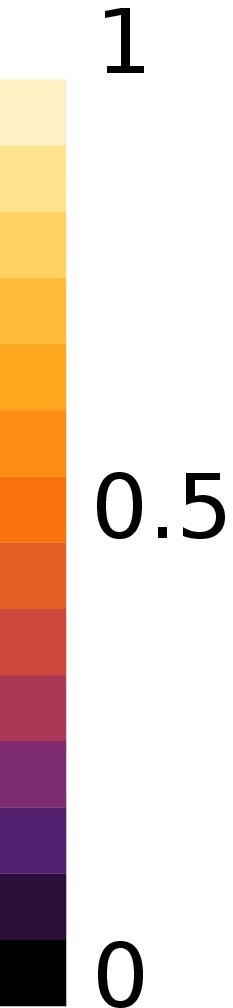}\\
		\end{tabbing}
		\caption{Theoretical plots of the degree of polarisation (colour online). We plot the state $\ket{\psi(\xi_H(r_H),\xi_V(r_V))}$ (upper row), $\ket{\alpha_H}_H\otimes\ket{\psi(\xi_V(r_V))}_V$ (middle row) and $\hat{D}(\alpha_H)\hat{\mathcal{S}}(0.2)\ket{0}_H\otimes\hat{D}(\beta_V)\hat{\mathcal{S}}(0.6) \ket{0}_V$ (lower row). The left plots show $\psc$, whereas the right plots show $\mathbb{P}_1$. 
		}
	\label{fig:Ptwomodestate}
\end{figure}

Finally, we consider the generalised pure two-mode squeezed state
\beq
\hat{D}(\alpha_H)\hat{\mathcal{S}}(\xi_H)\ket{0}_H\otimes\hat{D}(\alpha_V)\hat{\mathcal{S}}(\xi_V)\ket{0}_V,
\label{eq:experimentalstate}
\eeq
and plot the degree of polarisation (both $\psc$ (left column) and $\mathbb{P}_1$ (right column)) in \reffig{Ptwomodestate} for three different states. In \reffig{Ptwomodestate} (a), a two-mode vacuum state ($\alpha_H=\alpha_V=0$) is illustrated for different squeezing degrees. Both measures exhibit zero polarisation degree for equal squeezing parameters whereas for different squeezing parameters, $\mathbb{P}_1$ gives lower values than $\psc$. If we now set $\xi_H =0$ and $\alpha_V=0$ (corresponding to a coherent state in the $H$-mode and a squeezed vacuum state in the $V$-mode), the behaviour of the two polarisation measures is very different as illustrated in \reffig{Ptwomodestate} (b). Finally, we plot the two-mode displaced squeezed state (with $\xi_H=0.2$ and $\xi_V=0.6$) in \reffig{Ptwomodestate} (c). The plot for the semi-classical measure once again illustrates its inappropriateness to be a good measure of polarisation: According to $\psc$, by displacing a squeezed state further away from the vacuum, the state becomes more unpolarised. This incorrect behaviour is not seen by the new measure.

\textit{Experimental realisation} -
Since $\hat{S}_0$ commutes with all other Stokes operators, the Stokes vectors per excitation manifold and thus $\mathbb{P}_1$ can be directly accessed by using a proper waveplate configuration, a polarising beam splitter and two photon number resolving detectors (PNRDs). Such detectors are currently capable of efficiently detecting more than 6 photons and due to the rapid progress in developing such detectors more advanced versions with increased optical power range might soon become available~\cite{fukuda}. For very high excitations, standard intensity detectors can be used \cite{bowen, heersink05, iskhakov09}.

\begin{figure}
	\centering
		\includegraphics[scale=.6]{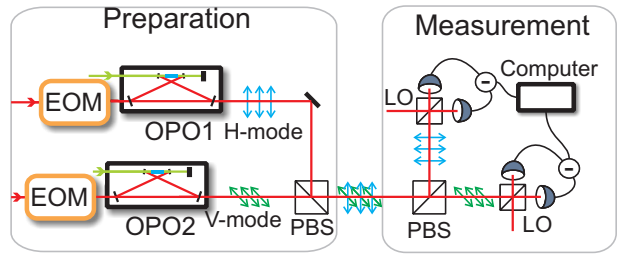}
		\caption{Setup for the production of the states in \refeq{experimentalstate}. A detailed description of the setup can be found in \cite{madsen}. For the results of this letter we don't use the squeezing of the second OPO since the interesting features of $\mathbb{P}_{1}$ can be experimentally shown without it (see text and \reffig{experimental_results2}).}
	\label{fig:setup}
\end{figure}

Since we wish to characterise the degree of polarisation in different regimes from low to high photon numbers, we have chosen to use a homodyne detector. Using such a detection device, a full tomographic reconstruction of the state \cite{smithey, vogel} from low to relatively high photon numbers is possible, and from this reconstruction we deduce the degree of polarisation.

We produce displaced two-mode squeezed states using the setup shown in \reffig{setup}. Two optical parametric oscillators (OPOs) based on nonlinear downconversion in periodically poled KTP crystals are used to generate vacuum squeezed states. The OPOs are injected with modulated coherent states to enable the production of displaced squeezed states~\cite{vertauschquetschverschiebe}. To form the two-mode state, the outputs from the OPOs are combined on a polarising beam splitter.

In contrast to previous realisations on CV polarisation quantum states, we solely define our state to be residing at a sideband frequency of 4.9 MHz \cite{andersen09}. Such a definition of the polarisation state enables us to investigate a large variety of different polarisation states from a low excitation to a relatively high excitation. We measure each mode, $H$ and $V$, by splitting the polarisation state on a polarising beam splitter and using two homodyne detectors. The measured currents of the homodyne detectors are sampled at 500 kHz with a frequency bandwidth of 90 kHz, and subsequently sent to a computer for analysis. Since the generated states have Gaussian wavefunctions, it suffices to estimate the covariance matrix of the state for full characterisation~\cite{weedbrook}. From this we calculate the first 50 excitation manifolds of the two-mode density matrix and take the expectation values of the Stokes operators (per manifold) from which the degree of polarisation is estimated.

\begin{figure}
		\includegraphics[scale=.55]{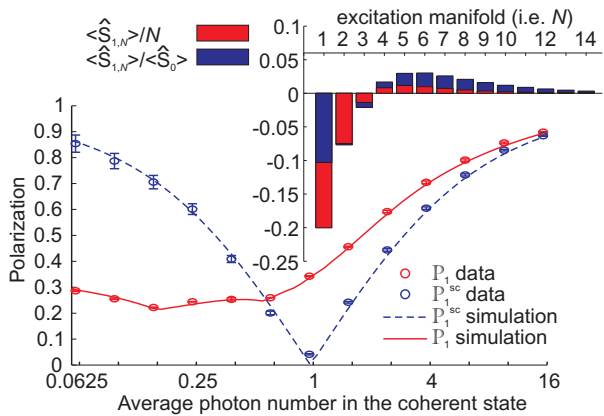}
		\caption{Degree of polarisation for experimental data of a squeezed state in mode $H$ and a displaced coherent state in mode $V$. We vary the displacement from $-6$ dB ($\alpha=0.25$, 0.0625 photons) to 6 dB ($\alpha=4$, 16 photons). The squeezed state has a squeezing of 3.2 dB and an antisqueezing of 7.4 dB, corresponding to 1.0 photons. Error bars correspond to 1\% uncertainty in the shot noise. Simulations starting from the initial squeezed state are shown with the solid lines. The inset shows the polarisation contributions of the different manifolds for the 0 dB (1 photon) state. Only $\langle\hat{S}_3\rangle$ contributes to the polarisation of the states produced here (i.e., $\langle\hat{S}_1\rangle=\langle\hat{S}_2\rangle=0$) and therefore one has $\psc=|\sum \mathrm{blue\; bars}|$ and $\mathbb{P}_1=\sum |\mathrm{red\; bars}|$.}
	\label{fig:experimental_results2}
\end{figure}

We start our experimental analysis with one-mode squeezed or one-mode coherent states as defined in \refeq{coherentstatedegofpol} and \refeq{onemodegeneral} (when $\alpha=0$), respectively. These states are produced by blocking OPO2 while operating either the EOM (for producing the coherent state) or the OPO1 (for producing the squeezed states).  The excitation of the coherent state is controlled by the modulation depth of the EOM whereas the squeezing degree (or the average number of photons associated with the squeezing process) of the squeezed state is controlled by the pump power. Our results for $\mathbb{P}_1$ and $\psc$ are plotted in \reffig{Ponemodestate}, where the error bars indicate the 1\% uncertainty in determining the shot-noise limit.  The experimental values for the squeezed state deviates slightly from the theoretical prediction (\refeq{onemodegeneral} with $\alpha=0$) which is a consequence of the small impurity of the generated state. As also predicted by theory, we see that both states become increasingly more polarised as the photon numbers from the coherent state or from the squeezed state increases.

Next, we investigate another particularly interesting state in which a coherent state is excited in the $H$-mode while the $V$-mode is a squeezed vacuum state corresponding to $\alpha_V=0$ and $r_H=0$. The squeezed state is squeezed by 3.2 dB below the shot noise limit and the coherent excitation of the $H$ mode is varied.

We present the experimental results for this state in \reffig{experimental_results2}. For a coherent amplitude of 0.25, $\psc$ yields a large degree of polarisation of 0.88 although this state is very close to the vacuum state. Furthermore, when increasing the coherent modulation, $\psc$ decreases to zero which occurs when the number of photons in each polarisation mode is unity. This result is erroneous as the state is not invariant to rotation (permutation of the H and V mode) for any value of the displacement. In contrast, the new measure is behaving as expected: The degree of polarisation is reasonably small for low excitations and increases near monotonically for larger excitations. These different behaviours can be understood by looking at the contributions of the different manifolds in definition (\ref{eq:new degree}). $\hat{S}_1$ is the only operator contributing to the polarisation and we plot the expectation value of this per manifold in inset of \reffig{experimental_results2}. Here, we see that it points in opposite directions for the different manifolds
which then sum up to zero for the $\psc$, and thus the polarisation becomes hidden. However, for the $\mathbb{P}_1$-measure, the polarisation is not hidden since in this case the absolute value of the $\langle \hat{S}_1\rangle$-values from the different manifolds are added.

 As a final experiment we operate both OPOs and modulators in order to produce the generalised state in \refeq{experimentalstate}. Also for these generalised states we measure a degree of polarisation which is monotonically increasing as a function of the coherent excitation.

\textit{Conclusion} - We have proposed a measure of a states first order polarisation (using the first moments of the Stokes operators) that overcomes the fallacities of the conventional measure. Specifically,  in contrast to the conventional measure, it detects first order hidden polarisation and it is continuous. Due to its suitability in quantifying polarisation and its extraordinary simplicity - can be directly measured - we believe that this measure of polarisation will have wide applicability in different sciences.

We acknowledges support from the Danish Research Council (Sapere Aude project no. 10-081599). GB acknowledges financial support from the Swedish Research Council (VR) through its Linn\ae us Center of Excellence ADOPT and contract 319-2010-7332. We acknowledge discussions with Jonas S\"{o}derholm.

\bibliography{shorttitles,polarisationcv18bib}

\begin{thebibliography}{10}
\bibitem{firstorder}%
  \BibitemOpen
  \bibinfo {note} {The subscript indicates that it only takes into account the
  first moments of the Stokes vectors. That is, for this paper we disregard all
  higher Stokes operator moments $\langle \hat{S}_j^r \rangle$, where $r>1$.
  Higher order of Stokes operaters have been investigated in \cite{klimov10}
  (second moments) and \cite{bjoerk12} (higher moments).}%
  \bibAnnoteFile{Stop}{firstorder}%
\bibitem{vertauschquetschverschiebe}%
  \BibitemOpen
  \bibinfo {note} {In practice we first displace and subsequently squeeze.
  However, these two operations are interchangeable when replacing
  $\alpha_{H,V}$ by
  $\alpha^\prime_{H,V}=\alpha_{H,V}\cosh(r_{H,V})+\alpha^\ast_{H,V}
  \exp(i\theta_{H,V})\sinh(r_{H,V})$ \cite{moeller, loudon}, and we choose the
  notation of \refeq{experimentalstate} to be consistent with the order in
  \refeq{onemodegeneral}.}%
  \bibAnnoteFile{Stop}{vertauschquetschverschiebe}%
\end{thebibliography}


\end{document}